\documentclass[12pt, letterpaper]{article}
\usepackage[margin=1in]{geometry}
\usepackage{graphicx} 
\usepackage{booktabs}
\usepackage{amsmath, amssymb}
\usepackage{arydshln}
\usepackage[utf8]{inputenc}
\usepackage{amsmath,amssymb,amsfonts}
\usepackage{xcolor}
\usepackage[authoryear]{natbib}
\setcitestyle{aysep={}}
\usepackage[table]{xcolor}
\usepackage{authblk}

\usepackage{setspace}
\usepackage{listings}
\usepackage{tikz}
\usepackage{hyperref}
\usepackage{dsfont}
\usepackage{bm}
\usepackage{float}
\usepackage{breakcites}
\usepackage{adjustbox}
\usepackage{graphicx}
\usepackage{caption}
\usepackage{titlesec}

\titleformat{\section}
  {\normalfont\large\bfseries\centering}
  {\thesection}
  {1em}
  {\MakeUppercase}

\title{
{\small\bfseries DRAFT}\\[0.6em]
Movement based social interactions with label uncertainty
}

\author[1]{Anisha J. Mathias}
\author[2]{John W. Durban}
\author[3]{Holly Fearnbach}
\author[3]{Samantha G. M. Leander}
\author[1]{Henry R. Scharf}

\affil[1]{Department of Mathematics, University of Arizona, 
    \newline 617 N. Santa Rita Avenue, Tucson, AZ 85721, USA}
\affil[2]{Southall Environmental Associates, Inc., 
    \newline 9099 Soquel Drive, Aptos, CA 95003, USA}
\affil[3]{SR3 SeaLife Response, Rehabilitation and Research, 
    \newline 2003 S. 216th St. \#98811, Des Moines, WA, 98198, USA}

\date{}

\begin{document}

\maketitle
\begin{abstract}
   Drone-based imagery provides high resolution data for tracking wildlife populations. However, monitoring marine mammals like cetaceans is challenging because individuals frequently submerge into water or move out of the drone’s field of view, resulting in disconnected trajectories due to uncertain identities. Existing social movement models assume individual identities are  known, which is often only true when data come from physical tracking devices like GPS collars. To address this limitation, we propose a Bayesian hierarchical state-space model that treats the identities associated with observed locations as unknown random variables. This allows us to jointly estimate individual movement parameters while propagating the uncertainty about identity. We apply our framework to drone-measured movement data from Risso’s dolphins (\emph{Grampus griseus}) to estimate the effects of sonar exposure on the dolphins' movement behavior. Our proposed method offers a solution for analyzing collective animal behavior when continuous tracking of every individual is challenging.

\end{abstract}

\section{Introduction} \label{sec:intro}
Advances in monitoring technologies have greatly improved the collection of information on wildlife populations. New instruments such as drones provide ways to track and observe animals and their environments with high resolutions remotely, generating information on movement, behavior, physiology, and habitat conditions. These tracking data allow researchers to address fundamental questions in movement ecology, including where animals are located, how they move through space, and which behavioral state they occupy \citep{dpvafuas, dbhrtav, qmbec}. However, for animals that travel in groups, movement decisions may depend not only on individual behavior but also on interactions among group members. The joint analysis of individual movement trajectories may therefore provide information about the strength of social relationships. Research in movement ecology has thus expanded to the study of coordinated movement among multiple animals, with increasing interest in whether spatial proximity, attraction, and directional coordination among animals reveal underlying social relationships \citep{eldmagm, dsnm,  irsimc}. 

  Attraction and alignment are the two social interaction parameters motivating our study, both of which are directly related to the social structure \citep{dsnmag}.  Attraction represents an individual’s tendency to move towards the mean position of connected individuals, whereas alignment represents an individual’s tendency to move in parallel with the trajectories of connected individuals. Drone data provide us with observations of individuals over time in the form of image sequences from which coordinate locations are derived. Many existing studies of social interactions among moving individuals rely on models that assume individual identities are fixed and known throughout the study period \citep{mgdam, jemibsiamd,  ammfmi, mifmia, mgbsic}. Assuming fixed and known identities is reasonable when individuals are uniquely identified and tracked consistently throughout the study period using tracking devices such as GPS collars, satellite, radio and acoustic tags.  However, with drone-based observations, challenges often arise when individuals are not easily distinguishable while moving in groups.  This issue is particularly acute for animals that move fast, and have high visual similarity or move in close proximity, making individual tracking difficult. \citet{amsqi} developed a multitracking system that tracked fish identity using their trajectories but found that identity tracking was unreliable after occlusion and separation events. Several other tracking tools have been developed to reduce identity ambiguity during animal interactions by using visual information to recognize individuals, reconnect trajectory segments after occlusions, and integrate pose estimation with individual identification and tracking \citep{idtiga, taeatowtma, mape}. These associated papers largely focus on either evaluating the tracking tools or estimating the behavioral parameters after reconstructing the trajectories,  with the identities treated as fixed and known.

  Our study focuses on the group movements of Risso's dolphins \emph{(Grampus griseus)}, a highly social species frequently observed moving in close-knit groups \citep{srd, irapsn}. This research is motivated by a dataset of Risso's dolphins from the Catalina Islands in Southern California, collected during a controlled exposure experiment \citep{irsmdce,brcdns}. Because these dolphins are commonly found in near-coastal navy training areas where mid-frequency active sonar is used, they are a relevant species for studying changes in social movement behavior under sonar exposure. Experts have studied  how whales and dolphins change their behavior when they are exposed to military sonar sounds using statistical models \citep{bwrsans, ammhmb, nbwipers, brstbbw}.  However, in our study, tracking individual dolphin trajectories remains challenging. Because they lack unique markings, animals that disappear from a drone's view for extended periods are often relabeled as entirely new individuals when they reappear. \citet{dynsn} addresses the problem of multi-labeling by constructing a proxy likelihood based on individuals observed at pairs of consecutive time steps, assuming that this subset is representative of the full population. Using this proxy likelihood, they approximately estimate social movement and network parameters of individuals moving in groups with uncertain identities.

 In our paper, rather than treating individual identities as fixed and known, we extend the dynamic social network movement models of \citep{dsnm,dynsn} by treating identity assignments as unknown random variables.  A closely related statistical approach is the spatial partial identity model (SPIM) \citep{scrpi}, which treats individual identity as unknown and uses spatial information together with partial identifying features to probabilistically link observations belonging to the same individual. In their ocelot application, left and right flank photographs provided partial identity information, and the model estimated posterior probabilities that observations from opposite side flanks  corresponded to the same individual. While this approach directly addresses identity uncertainty, the SPIM framework is focused on spatial capture-recapture rather than continuous group movement. In contrast, we jointly infer identity assignments and social movement parameters from group movement trajectories observed at one-second discrete time intervals, thereby propagating identity uncertainty into inference on the movement interaction parameters. 
 
 The remainder of this paper is organized as follows. Section (\ref{sec:method}) presents the state-space model with
time-varying parameters to capture social movement dynamics and introduces generalizations to allow for identity uncertainty. The continuous movement process is modeled using a linear, Gaussian state-space formulation, which allows for likelihood evaluation through Kalman filter recursions.  Section (\ref{sec:implementation}) describes the implementation strategy used for inference. We demonstrate the proposed model for the joint estimation of unknown identities and behavioral components through a simulation study in section (\ref{sec:simulation_study}) and  apply the same model to Risso's dolphins in section (\ref{sec:application}). Finally, we close this paper with the discussion of the approach and future directions.

\section{methods} \label{sec:method}
We propose a Bayesian hierarchical state-space model (SSM) for jointly modeling animal movement and identity uncertainty, adapted from previous movement research \citep{dsnm, dynsn}. The model combines a continuous-space, discrete-time state-space formulation with a Gaussian Markov random field (GMRF) to capture social dependence among individuals. This formulation is particularly well suited to our collective movement data, where locations are observed at regular time points and movement decisions may be influenced by neighboring group members. 

At any given time $t$, the drone's field of view captures only a subset of size $n_{t}$ of the total, known population of size $J$. Let ${\mu}_{it}$ denote the true position of an individual $i$ at time $t$. Let $\bm{\mathrm{\mathit{\mu}}}_t=(\mu_{1t}, \mu_{2t}, ...,\mu_{Jt})'$ be a  column vector containing the underlying true positions of all $J$ individuals at $t$.  The observed positions of the $n_t$ individuals at time $t$ are written as $\bm{\mathrm{y}}_t$, where $\bm{\mathrm{y}}_t$ is a length-$n_t$ column vector. 

 The linear Gaussian SSM is a dynamic stochastic system,  consisting of an unobserved Markov state process and an observed data process, defined by 
\begin{equation}
    \mathrm{\bm{\mu}}_{t+1}^{(r)} = \bm{\mathrm{A}}_{t}\mathrm{\bm{\mu}}_{t}^{(r)} + \bm{\mathrm{\eta}}_t^{(r)}, \qquad \bm{\mathrm{\eta}}_t^{(r)} \sim \mathrm{N}(0, \mathrm{\sigma}^2\bm{\mathrm{Q}}^{-1}_{t})
    \label{process_equation}
\end{equation}
\begin{equation}
\bm{\mathrm{y}}_t^{(r)}= \bm{\mathrm{Z}}_t\mathrm{\bm{\mu}}_t^{(r)} + \bm{\mathrm{\varepsilon}}_t^{(r)}, \qquad \bm{\mathrm{\varepsilon}}_t^{(r)} \sim \mathrm{N}(0,\mathrm{\tau}^2\bm{\mathrm{I}})
\label{obs_equation}
\end{equation}
where $t=1,\dots,T$. Let $ r \in \{ x_1, x_2 \} $, where $x_1$ and $x_2$ represent the spatial coordinate indices in the real plane. Spatial coordinates associated with each index are modeled independently using two identical SSMs. Therefore, from this point onward, unless stated explicitly, we use notation for a single coordinate.  

The process model for the SSM, given in equation (\ref{process_equation}), is formulated as a correlated random walk describing the temporal evolution of the underlying true positions of the animals at $t$. The observation model, given in equation (\ref{obs_equation}), defines the conditional link between the observed data and the underlying true positions generated by the process model. The measurement error vector $\bm{\mathit{\varepsilon}}_t$ is assumed to be normally distributed with mean zero  and $\tau^2$ is the scalar measurement variance, which represents the imprecision of the drone-based location observations. We expect high spatial precision in the observations, so we assume $\tau^2$ to be small and known throughout the observation period.

\subsection{Behavioral components}
As in \cite{dsnm}, we define social relations among the population using  dynamic, binary adjacency matrices $\bm{\mathrm{W}}_t$ indexed at discrete time steps, where $w_{ijt}= 1$ indicates a social relation among individuals $i,j \in \{1,\ldots,J \}$ at time $t$.  We define $\overline{\bm{\mathrm{W}}}_t$ as the row-normalized adjacency matrix, with elements 
\begin{equation}
    \overline{w_{ijt}}= \frac{w_{ijt}}{w_{i^{+}t}}, 
\end{equation}
where $w_{iit} =1$ and $w_{i^{+}t} = \sum_{j=1}^{J} w_{ijt}$ are the number of socially connected individuals with an individual $i$ at time $t$. For a given social structure, individual movement can be characterized by several social behaviors. We focus on two primary behaviors: attraction, denoted by $\mathrm{\beta}_t $, which describes the individual's movement towards the average position of socially connected individuals, and alignment, denoted by $\mathrm{\alpha}_t $, which describes the coordinated movement in parallel with those individuals. We explain the role of $\beta_t$ and $\alpha_t$ in this section below.

The transition matrix $\bm{\mathrm{A}}_t$ is defined to be
\begin{equation}
    \bm{\mathrm{A}}_t = (1-\beta_t)\bm{\mathrm{I}}
    + \beta_t \overline{\bm{\mathrm{W}}}_t.
    \label{transition}
\end{equation}
This matrix determines the expected positions of individuals,
$\mathrm{E}(\boldsymbol\mu_{t+1} \mid \boldsymbol\mu_t)$,
at time $t+1$, conditional on their positions at time $t$.
The parameter $\beta_t \in [0,1]$ controls the extent to which
an individual's expected position is pulled toward the mean
position of its socially connected individuals.

 When values of $\beta_t$ are near $0$, the expected position is centered at the individual's previous position, and there is no attraction. As $\beta_t$ approaches $1$, an individual is expected to move towards the average position of all socially connected individuals, including itself. On average, an individual closes a proportion $\beta_t$ of the full distance between its current location and the mean location of its connected individuals during a single time step. 

The transition matrix $\bm{\mathrm{A}}_t$ defines the expected movement from one time point to another. The vector $\bm{\eta}_t$ in equation (\ref{process_equation}) represents correlated variation around the expected locations, where $\sigma^2$ controls the overall magnitude of that variability.  The precision matrix $\bm{\mathrm{Q}}_t$ determines the covariance structure of that movement variability among individuals. In a GMRF, the precision matrix is useful because its structure directly represents conditional dependence among individuals. Specifically, a zero off-diagonal element in the precision matrix means the two individuals are conditionally independent, given all other individuals in the model. Conversely, a non-zero off-diagonal element signifies a conditional dependence, the strength of which is controlled by the alignment parameter $\alpha _{t} \in [0,1]$. The elements of the precision matrix $\bm{\mathrm{Q}}_t$ at time $t$ are
 \begin{equation}
     \mathrm{Q}_{ijt} =
\begin{cases}
-\alpha_t w_{ijt}, & i \neq j \\[1mm]
w_{i^+t}, & i = j
\end{cases}.
 \end{equation}
 Values of $\alpha_t$ near $0$ correspond to little to no alignment, while values near $1$ correspond to  perfect alignment, where an individual moves in parallel with its connected individuals.  We assume the movement effects are the same for all individuals in the population. Thus, the coefficients $\mathrm{\beta}_t$ and $\mathrm{\alpha}_t$ represent population-level attraction and alignment parameters. 

To incorporate the effects of environmental variables on behavior, we follow the same approach implemented by \cite{dynsn}, who apply logit linear to models for $\alpha_t$, and $\beta_t$. This allows the strength of attraction and alignment to vary in response to natural and anthropogenic covariates. We apply this model to the behavioral effects, such that

\begin{equation}
\mathrm{logit}(\alpha_t)=\bm{\mathrm{x}}_t'\boldsymbol{\gamma}_\alpha,
\qquad
\mathrm{logit}(\beta_t)=\bm{\mathrm{x}}_t'\boldsymbol{\gamma}_\beta,
\label{logit_link}
\end{equation}
where the row vector $\bm{\mathrm{x}}_t' = (x_{0t}, x_{1t}, x_{2t})$
represents time-indexed covariates of interest. In our application,
these covariates represent the different phases of the naval sonar
exposure experiment. The intercept $x_{0t}=1$ at all time points,
while $x_{1t}=1$ during sonar exposure and 0 otherwise, and
$x_{2t}=1$ during the post-exposure period and 0 otherwise.
Thus, the pre-exposure period serves as the baseline, when
$x_{1t}=x_{2t}=0$. This formulation allows $\alpha_t$ and
$\beta_t$ to vary among the pre-exposure, exposure, and
post-exposure periods. The regression coefficients
$\bm{\mathrm{\gamma}}_{\alpha}=
(\mathrm{\gamma}_{\alpha_0}, \mathrm{\gamma}_{\alpha_1},
\mathrm{\gamma}_{\alpha_2})'$
and
$\bm{\mathrm{\gamma}}_{\beta}=
(\mathrm{\gamma}_{\beta_0}, \mathrm{\gamma}_{\beta_1},
\mathrm{\gamma}_{\beta_2})'$
quantify the baseline social behavior and changes in the
tendency of individuals to align with and move toward
their socially connected individuals across
the three exposure periods.

\subsection{Modeling Identity Uncertainty}\label{sec:identity_assignment}
 We define $\bm{\mathrm{Z}}_t$ in the observation equation (\ref{obs_equation}) as an $n_t \times J$ binary matrix for every time $t$. For each $i=1,2,\dots,n_t$, a single non-zero entry $z_{ijt} = 1$ in row $i$ of $\bm{\mathrm{Z}}_t$ indicates that observation $y_{it}$ at time $t$ corresponds to individual $j$. Since each non-zero entry $z_{ijt}$ corresponds to a unique observation $y_{it}$, the column sums of $\bm{\mathrm{Z}}_{t}$ must either be equal to  $0$ or $1$ to ensure that each observation at time $t$ maps to exactly one individual from $J$.
  For drone-based observations, individuals cannot always be recognized with certainty across frames and thus $\bm{\mathrm{Z}}_{t}$ is partially unknown. While a high frame-rate often allows for reliable individual tracking across consecutive frames, identity uncertainty arises when an animal moves out of the field of view and subsequently reappears. Our application dataset tracks observed individuals at time  $t$ using timestamped, spatial coordinates, and labels. In this dataset, individuals observed at time $t+1$ who could not be confidently linked to an individual observed at time $t$ were treated as newly observed individuals and assigned new labels. Thus, the observed locations of a single individual are in general assigned multiple labels. 

To account for identity uncertainty, we model the entries of the identity assignment matrix, $\bm{\mathrm{Z}}_{t}$ as unknown random variables. Without loss of generality, assume that the $n_1$ individuals observed at the first time step are assigned labels $1,\ldots,n_1$ and thus $\bm{\mathrm{Z}}_1 =\begin{bmatrix} \bm{\mathrm{I}}_{n_1} & \bm{\mathrm{0}}_{n_1 \times (J - n_1)} \end{bmatrix} $. We specify a first order Markov process in discrete time $t$ with $t \geq 2$,  
\begin{equation}
    \bm{\mathrm{z}}_{1t} \mid \bm{\mathrm{Z}}_{t-1} \sim \mathrm{Multinomial} (\bm{\mathrm{\pi}}_{1t})
    \label{Z_eq}
\end{equation}
\begin{equation}
  \bm{\mathrm{z}}_{it}  \mid \bm{\mathrm{Z}}_{t-1}, \bm{\mathrm{z}}_{1t}, \bm{\mathrm{z}}_{2t}, \dots, \bm{\mathrm{z}}_{(i-1)t} \sim \mathrm{Multinomial}(\bm{\mathrm{\pi}}_{it}), \quad i= 2,\dots,n_t
  \label{Z_equation}
\end{equation}
where $\bm{\mathrm{z}}_{it}$  is a row vector linking the underlying identities  to the observation $y_{it}$. Here, $\bm{\mathrm{z}}_{1t}, \bm{\mathrm{z}}_{2t}, \dots, \bm{\mathrm{z}}_{(i-1)t}$,
also written as $\bm{\mathrm{z}}_{<it}$, represent the identities for the observations that have already been assigned at time $t$, before observation $i$. The probability vector $\bm{\mathrm{\pi}}_{it} = (\pi_{i1t}, \pi_{i2t}, \dots, \pi_{iJt})$ with $\sum_{j=1}^{J} \mathit{\pi}_{ijt} = 1$, contains the conditional probabilities of assignment to each possible individual. Its element $\pi_{ijt} = P(z_{ijt}=1 \mid \bm{\mathrm
{Z}}_{t-1}, \bm{\mathrm{z}}_{1t}, \ldots, \bm{\mathrm{z}}_{(i-1)t})$ represents the probability that the observation $y_{it}$ at time $t$ corresponds to the individual $j$, conditioned on the assignments from the previous time step and the preceding assignments at the current time step.

Let $\mathcal{L}_t = \{L_{1t}, \dots, L_{n_t t}\}$ be the set of labels assigned to the $n_{t}$ individuals observed at time $t$ in the dataset.  If, for example, time gaps between observations are large,  individuals may not be recognizable across consecutive time steps using spatial coordinates alone. In this case, the recorded observations would be assigned labels $\mathcal{L}_t$ such that $\mathcal{L}_t \cap \mathcal{L}_{s} = \emptyset \ \forall \ t \neq s$ where $ t,s \in \{1,\ldots,T\}$. Consequently, the total number of unique labels would equal the total number of observations. In this setting, the conditional probability $\pi_{ijt}$ is independent of $\bm{\mathrm{Z}}_{t-1}$. 

\subsubsection{Constraints on identity assignments}
In our application, observations are recorded at a high-temporal resolution of one-second intervals, providing the capacity to recognize individuals at consecutive times, even though an individual's true identity may not be known. Identical labels across consecutive timestamps indicate that a specific individual is recognizable. 

Let $\mathcal{R}_{(t, t+1)} = \left \{ \left (k,i \right ) : L_{kt} = L_{i(t+1)}, k \in \{1,\ldots,n_t\}, i \in \{1,\ldots,n_{t+1}\} \right \}$ denote the set of pairs of row indices whose labels in the dataset match at times $t$ and $t+1$. Recognition imposes constraints on some rows of $\bm{\mathrm{Z}}_{t+1}$, by collapsing the conditional distribution of $\bm{\mathrm{z}}_{i(t+1)}$ to the recognized individual at the previous time point, 

\begin{equation}
   P(\bm{\mathrm{z}}_{i(t+1)} = \bm{\mathrm{z}}_{kt}|\bm{\mathrm{Z}}_t, \bm{\mathrm{z}}_{<i(t+1)}) = 1, \quad (k,i) \in \mathcal{R}_{(t, t+1)}.
\end{equation}
In principle, an individual may be recognized after a longer observation gap if it can be uniquely identified from distinctive features. In such cases, observations recorded at two non-consecutive time points may receive the same observed label, and the recognition set can be generalized to $\mathcal{R}_{(t,t^*)}$ for any $t<t^*$. Our current formulation constructs the identity assignments based on current and  previous time point assignments. However, equation (\ref{Z_equation}) could be extended so that the identity assignment at time $t^*$ depends on all the previous assignments until time $t$ rather than only on the immediately preceding time point. Thus, if an observed label at $t^*$ matches a label previously observed at time $t$, the corresponding rows of $\bm{\mathrm{Z}}_{t^*}$ and $\bm{\mathrm{Z}}_t$ could be constrained to have the same identity assignments.

In addition to the certainty about the recognition of some individuals, the dataset labels provide information about some of the obscured individuals at a given time step. Let $\mathcal{P}_t=\left \{k\in\left \{1,\ldots, n_t \right \}: L_{kt} \in L_{t}\backslash L_{t+1} \right \}$ denote the set of row indices at time $t$ whose observed labels are not used at time $t+1$. These provide an exclusion constraint on the possible identity assignments in $\bm{\mathrm{Z}}_{t+1}$. Then, for every $i = 1, \ldots, n_{t+1}$, 
\begin{equation}
   P(\bm{\mathrm{z}}_{i(t+1)} = \bm{\mathrm{z}}_{kt}|\bm{\mathrm{Z}}_t, \bm{\mathrm{z}}_{<i(t+1)}) = 0, \quad k\in \mathcal{P}_t.
\end{equation}
We assume this constraint because an individual seen at consecutive time points would have been recognized while the observations were recorded.  Thus, each label used at $t$ but not at $t+1$ implies that the corresponding individual was outside the field of view and therefore not one of the $n_{t+1}$ individuals observed at time $t+1$.

Conditioned on the assignments from previous time points and subject to the constraints implied by the labels, some rows of $\bm{\mathrm{Z}}_{t}$ become conditionally known, while the remaining rows are modeled as random vectors. Because we have no prior distinguishing information about individuals in the population, we place a uniform prior across all assignments respecting constraints implied by the labels, independent of the duration an individual has remained out of sight.

\begin{figure}[H]
\hspace{-4mm}
    \includegraphics[width=1.03\textwidth]{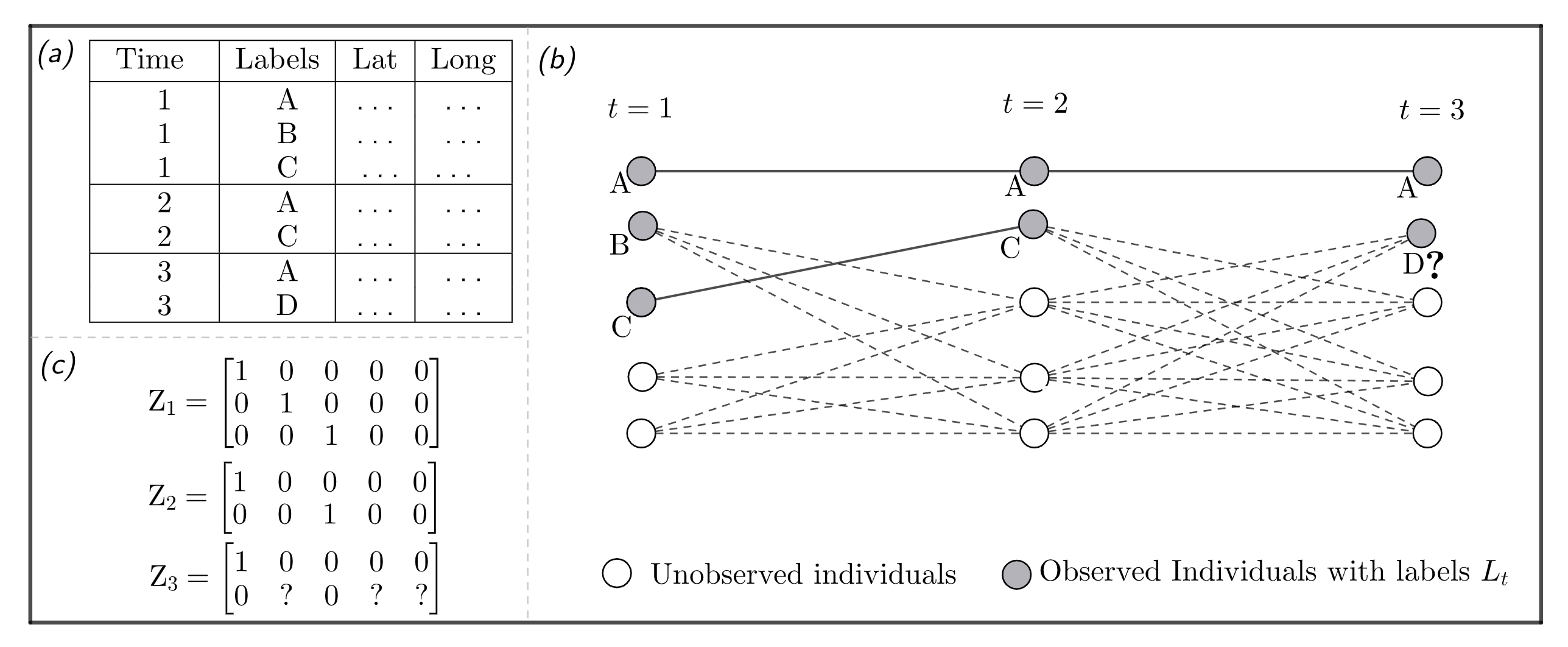}
    \caption{ Labeled dataset and identity uncertainty. (a) Sample dataset for $J=5$ individuals, over $3$ time steps. (b) Schematic representation of individual trajectories within the population. Bold lines track the movements of recognized individuals, while dotted lines denote gaps in tracking caused by out of sight periods of the individuals. (c) Identity assignment matrices for three time steps.
 }
    \label{identity_assignment}
\end{figure}

We use an example to illustrate the identity model and constraints encoded by recorded labels. Figure (\ref{identity_assignment}a) shows a sample dataset for observed individuals with recorded labels and Figure (\ref{identity_assignment}b)  shows a diagram of these data with known population size $J=5$. This example matches how our application assigns labels to identities. At $t=1$, there are three individuals in the population with assigned labels $\{A, \ B , \ C \}$  and thus the identity assignment matrix is $\bm{\mathrm{Z}}_1$= $\begin{bmatrix} \bm{\mathrm{I}}_{3} & \bm{\mathrm{0}}_{3 \times 2} \end{bmatrix}$, see Figure  (\ref{identity_assignment}c). At time $t=2$, both the observed individuals assigned labels $A$ and $C$ are recognizable, and thus conditionally known. At $t=3$, the individual with label $A$ remains recognizable, generating the probability assignment vector $\bm{\mathrm{\pi}}_{13} = (1, 0, 0, 0, 0)$.  In Figure (\ref{identity_assignment}a), the recorded label in the dataset corresponding to the second observation is $D$ at time $t=3$ because it was implicitly determined not to be either $A$ or $C$. Thus, this observation could correspond to the individual assigned label $B$ (last seen at $t=1$), or to one of the two remaining unobserved individuals in the population.  Conditional on the assignments at $t=2$ and the preceding assignments at $t=3$,  the probability assignment vector is $\bm{\mathrm{\pi}}_{23} = (0, \frac{1}{3}, 0, \frac{1}{3}, \frac{1}{3})$.

The main methodological contribution of this work is a state-space movement model that explicitly accounts for uncertainty in individual identity. Unlike standard approaches that assume observations are correctly assigned to individuals, our framework treats the identity assignments as unknown and propagates this uncertainty through the movement model. This allows inference on movement parameters to reflect uncertainty in which observations belong to which individuals, reducing the risk of biased conclusions arising from mistaken identities.  Inference under the discrete-time hierarchical state space movement model, as stated in equation (\ref{process_equation}) and equation (\ref{obs_equation}), has three sets of unknown quantities: $(1)$ the underlying true movement trajectories $\bm{\mathrm{\mu}}_{1:T} = \{\bm{\mathrm{\mu}}_{1}, \bm{\mathrm{\mu}}_{2}, \dots, \bm{\mathrm{\mu}}_{T}\}$, $(2)$ the latent behavioral parameters governing the movement $\bm{\theta} = $\{$\bm{\gamma}_{\alpha}$, 
$\bm{\gamma}_{\beta}$, $\sigma^2\} $ and $(3)$ the identity assignment matrices $\bm{\mathrm{Z}}_{1:T}$. In our application, the drone-based data provide only the observed locations of individuals at each time point. The recorded labels provide partial information about identity continuity, allowing some entries in the $\bm{\mathrm{Z}}_t$ matrices to be treated as conditionally fixed assignments when the same label is observed continuously across adjacent time points. Additionally, observations associated with individuals that were not observed at the previous time point are represented through rows of $\bm{\mathrm{Z}}_t$ with non-trivial probability distribution.

\section{Implementation}\label{sec:implementation}
 While it is natural in some applications to treat social networks $\bm{\mathrm{W}}_{1:T}$ as unknown, in our application we assume this binary network to be fully connected and static. The fully connected structure $\mathrm{w}_{ijt} = 1 $ represents the assumption that all group members influence one another. We define the collection of observation vectors through time $t$ as  $\bm{\mathrm{y}}_{1:t} = \{\bm{\mathrm{y}}_1, \bm{\mathrm{y}}_2, \ldots,\bm{\mathrm{y}}_t\}$. Bayesian inference is based on the joint distribution of $\bm{\bm{\theta}}$, $\bm{\mathrm{Z}}_{1:T}$, and $\bm{\mu}_{1:T}$ conditional on the observed data $\bm{\mathrm{y}}_{1:T}$, which we approximate using samples from the joint posterior. Sampling directly from the joint posterior of $\bm{\bm{\theta}}$, $\bm{\mathrm{Z}}_{1:T}$ and the latent trajectories of all $J$ individuals is inefficient because of the high dimensionality of the state process. Thus, we integrate out the posterior distribution $\bm{\mathrm{\mu}}_{1:T}$ to obtain the marginal likelihood  $p(\bm{\mathrm{y}}_{1:T}\mid \bm{\bm{\theta}}, \bm{\mathrm{Z}}_{1:T})$ and target $p(\bm{\theta}, \bm{\mathrm{Z}}_{1:T} \mid \bm{\mathrm{y}}_{1:T})$. Since our model is a linear, Gaussian SSM this integration can be evaluated through Kalman filter recursions.

The Kalman filter is a recursive algorithm that moves forward through time, updating the predicted mean $\bm{\mathrm{\mathit{a}}}_t = E(\bm{\mathrm{\mu}}_{t}\mid \bm{\mathrm{y}}_{1:t-1})$ and covariance matrix $\bm{\mathrm{P}}_t = \mathrm{Var}(\bm{\mathrm{\mu}}_{t}\mid \bm{\mathrm{y}}_{1:t-1})$, where $\bm{\mathrm{\mathit{a}}}_t$  is the mean and $\bm{\mathrm{P}}_t$ is the variance of the underlying true positions $\bm{\mathrm{\mu}}_t$ before seeing the observation $\bm{\mathrm{y}}_t$ at time $t$. We assume that the initial state vector $\bm{\mathrm{\mu}}_1$ follows $\mathrm{N}(\bm{\mathrm{\mathit{a}}}_1, \bm{\mathrm{P}}_1)$ where $\bm{\mathrm{\mathit{a}}}_1$ and $ \bm{\mathrm{P}}_1$ denote the prior mean vector and covariance matrix of the initial state, respectively. The Kalman filter evaluates the marginal likelihood recursively, e.g. \citep{tsassm} and thus the likelihood is

\begin{equation}
         L(\bm{\mathrm{Z}}_{1:T}, \bm{\theta} \mid \bm{\mathrm{y}}_{1:T}) = p(\bm{\mathrm{y}}_{1:T}\mid \bm{\theta}, \bm{\mathrm{Z}}_{1:T}) = p(\bm{\mathrm{y}}_1)\prod_{t=2}^{T} p(\bm{\mathrm{y}}_t\mid \bm{\mathrm{y}}_{1:t-1},\bm{\mathrm{Z}}_{1:t},\bm{\theta}), 
      \label{likelihood}
\end{equation}
where $p(\bm{\mathrm{\mathit{y}}}_t\mid \bm{\mathrm{\mathit{y}}}_{1:t-1},\bm{\mathrm{Z}}_{1:t},\bm{\theta})$ = $\mathrm{N}\left(\bm{\mathrm{Z_{t}}}\bm{\mathrm{a_{t}}}, \bm{\mathrm{Z}}_{t} \bm{\mathrm{\mathit{P}}}_t \bm{\mathrm{Z}}_{t} ' + \tau^2\bm{\mathrm{I}}\right)$.

To complete our analysis, we now define the prior distributions over the model parameters. We will assign  priors for $\bm{\theta}$ based on the previously defined priors in the papers \citep{dynsn,  dsnm}. This allows us to compare some of the results from these papers to the new results in this paper. For the regression coefficients, we specify normal priors, and for the  variance parameter $\sigma^2$, we assign an inverse-gamma prior.

\subsection{Posterior sampling}

We use a Metropolis-Hastings within Gibbs algorithm to sample from the joint posterior distribution $p\left(
\bm{\theta},\bm{\mathrm Z}_{1:T}
\mid
\bm{\mathrm{y}}_{1:T}
\right)
$. At each MCMC iteration, the movement parameters in $\bm{\theta}$ are updated individually using an adaptively tuned random walk,
conditional on the latest values of identity assignment, the current values of other components in $\bm{\theta}$, and the observed data $\bm{\mathrm{y}}_{1:t}$. The corresponding multinomial variables $\bm{\mathrm{z}}_{it}$ are updated using a discrete multinomial random walk proposal. Identity assignments are updated sequentially, row by row. At each iteration, the sampler proposes moving the current identity assignment $z_{ijt}=1$ to another identity $z_{ij^*t}=1$, with equal probability for all $j^*$. The proposed assignment is then accepted or rejected according to the Metropolis algorithm based on its full conditional probability under the movement model.  The algorithm was implemented using the R package NIMBLE  \citep{nimble2026}, which allows users to specify a model and prior distributions that are used to automatically create an MCMC algorithm with the posterior as its stationary point.

\section{Simulation study} \label{sec:simulation_study}

We performed a simulation study to evaluate the ability of our proposed method to recover known movement parameters $\bm{\theta}$. We considered multiple scenarios based on varying durations of stable observed group composition before the set of visible individuals changed, representing their temporary absence and reappearance in the drone's field of view. Motivated by our application consisting of sequential phases of controlled exposure experiments (`pre-exposure', `exposure', and `post-exposure'), we construct covariates $\bm{\mathrm{x}}_t$ by dividing the entire study period, $T$, into three approximately equal intervals. In this timeline, the sonar begins at $t_{\mathrm{start}}$ and concludes at $t_{\mathrm{stop}}$. 

The duration of group stability varied across three scenarios, differing in how long the group composition remained the same before a change occurred. A change in the observed group composition at time $t+1$ was defined as either an individual becoming visible in the drone's field of view who was not visible in the frame at the preceding time point $t$, or an individual observed at time $t$ going out of sight at time $t+1$.  Let $G_t$ denote the set of individuals observed at time $t$. The variable $O_t \in \{0,1\}$ indicates whether this group $G_t$ can change between times $t$ and $t+1$, with $O_t = 1$ denoting a possible change and $O_t=0$ denoting no change. We model $O_t$ as a Bernoulli random variable with success probability $\rho = p(O_t=1)$, where $\rho \in \{0.05, 0.5, 0.75\}$. A lower change probability $\rho = 0.05\ $ indicates longer consecutive periods of group stability, allowing identities to be retained with certainty over more successive time points, and thus fewer rows $\bm{\mathrm{z}}_{it}$ correspond to uncertain identities.  Conversely, as $\rho$ increases to $0.50$ and $0.75$, the group composition changes more frequently. This adds more unknown random variables to $\bm{\mathrm{z}}_{it}$, expanding the dimension of the parameter space.

\begin{figure}[H]
    \centering
    \hspace{-4mm}
    \includegraphics[width=1.02\textwidth]{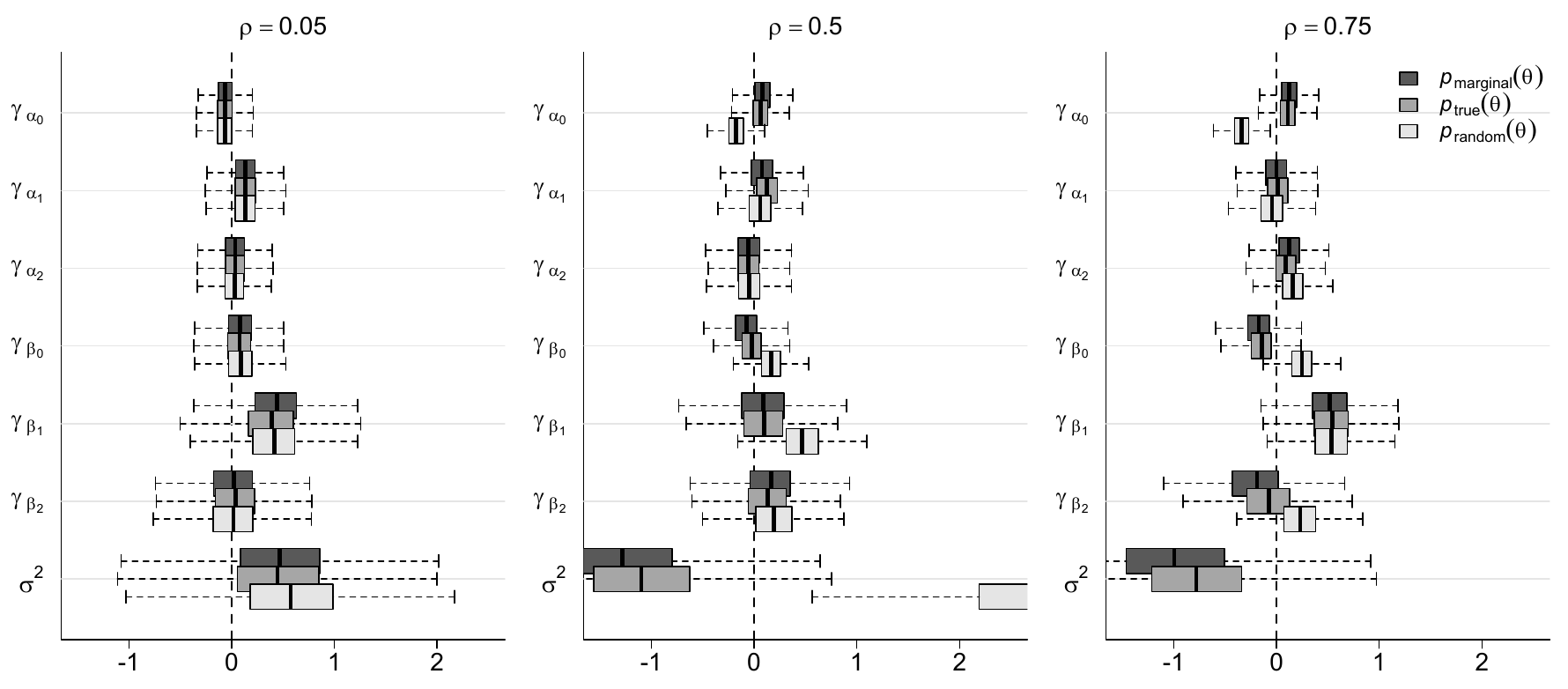}
            \vspace{2.5mm}
    \caption{Boxplots of the differences between the posterior samples and the parameter values used to generate the simulated data, for samples drawn from $p_{\mathrm{marginal}}(\bm{\theta})$, $p_{\mathrm{true}}(\bm{\theta})$, and $p_{\mathrm{random}}(\bm{\theta})$ under the three stability group composition scenarios defined by success probability $\rho\in\{0.05,0.50,0.75\}$ }

    \label{boxplot_figure_1}
\end{figure}

In each scenario, we simulated $J= 10$ individuals over an observation period of $T= 400$ time steps. To make the simulation study closely reflect the application study, we constructed the simulated datasets to preserve key features of the observed data. Conditioned on $O_{t}=1$, a possible change in group composition was initiated between times $t$ and $t+1$. The group $G_{t+1}$ was generated by randomly sampling individuals into three distinct subsets: individuals who remained observed across times $t$ and $t+1$, $G_t\cap G_{t+1}$; those who left the observed group after time $t$, $G_t\setminus G_{t+1}$, and became available to re-enter the observed population from time $t+2$ onward; and those drawn from the available population at time $t+1$ who became visible in the drone's frame at time $t+1$, $G_{t+1}\setminus G_t$.  For each value of $\rho$, we produced $10$ realizations of $\bm{\mathrm{Z}}_{1:T}$, with each realization used to simulate a corresponding dataset $\bm{\mathrm{y}}_{1:T}$. For each value of $\rho$, we present results from
one simulated dataset in Figure~\ref{boxplot_figure_1}.

To evaluate the performance of the model in estimating movement parameters, we simulated correlated random-walk movement data under fixed true values of the identity assignment matrices $\bm{\mathrm{Z}}_{1:T}$ and movement parameters $\bm{\theta}$, where $\gamma_{\alpha_{0}} = 4, \gamma_{\alpha_{1}} = -1.3, \gamma_{\alpha_{2}} = -1.5 $, $\gamma_{\beta_{0}} = -2, \gamma_{\beta_{1}} = -2, \gamma_{\beta_{2}} = -2 $ and $\sigma^2 =  16$. These coefficient values correspond to weak attraction $\beta_t$ and strong alignment $\alpha_t$. These were set based on the characteristics of the movement behavior of Risso's dolphins observed in the previous research paper, see \cite{dynsn}. We compared the posterior distributions of $\bm{\theta}$ under three model fits. For the first model, we jointly inferred $\bm{\theta}$ and the unknown identity assignments $\bm{\mathrm{Z}}_{1:T}$ to obtain the marginal posterior distribution $p_{\mathrm{marginal}}(\bm{\theta}) = p(\bm{\theta}|\bm{\mathrm{y}}_{1:T})$. This distribution accounted for the underlying uncertainty in $\bm{\mathrm{Z}}_{1:T}$. For the second model, we conditioned the posterior on the true identity assignment matrices used to generate the simulated dataset $\bm{\mathrm{y}}_{1:T}$, yielding $p_{\mathrm{true}}(\bm{\theta}) =  p(\bm{\theta}|\bm{\mathrm{y}}_{1:T}, \bm{\mathrm{Z}}_{1:T}^{\mathrm{true}})$. For the third model, we randomly assigned labels to rows with uncertain identities, filled the remaining rows according to the label imposed constraints, and fixed the resulting matrix $\bm{\mathrm{Z}}_{1:T}^{\mathrm{random}}$ to estimate the posterior distribution $p_{\mathrm{random}}(\bm{\theta}) =  p(\bm{\theta}|\bm{\mathrm{y}}_{1:T},\bm{\mathrm{Z}}_{1:T}^{\mathrm{random}})$.

Figure (\ref{boxplot_figure_1}) shows the marginal posterior distributions of each element of $\bm{\theta}$, minus the value used in simulation, for samples drawn from $p_{\mathrm{marginal}}(\bm{\theta})$, $p_{\mathrm{true}}(\bm{\theta})$, and $p_{\mathrm{random}}(\bm{\theta})$ under $\rho\in\{0.05,0.50,0.75\}$. Across all three scenarios, the discrepancies obtained from $p_{\mathrm{marginal}}(\bm{\theta})$ and $p_{\mathrm{true}}(\bm{\theta})$ remain similar for most parameters, indicating that accounting for identity uncertainty produces inference close to that obtained when the true identity assignments are known. In contrast, the posterior discrepancies from $p_{\mathrm{random}}(\bm{\theta})$ generally increase as $\rho$ increases, with particularly noticeable deviations for several attraction coefficients and for $\sigma^2$. For $\rho=0.75$, the discrepancy in $\sigma^2$ extends beyond the plotting range and is therefore not fully visible. Larger values of $\rho$ allow for more frequent changes in the observed group composition, resulting in a greater number of uncertain identity assignments being replaced by random fixed labels in the randomly labeled model fit. The posterior discrepancies under this model suggest that incorrectly fixing uncertain identities increasingly gives deviated inference as individual trajectories become more fragmented compared to when we propagate uncertainty into the model.

\section{Application}
\label{sec:application}
We fit our model to the data collected using drone data capturing the joint movement of Risso's dolphins observed near Santa Catalina Island in Southern California. The drone based photogrammetry allowed us to obtain high-resolution spatial and temporal observations of multiple individuals moving together. The dataset consisted of observed locations derived from images taken at one-second intervals for a total observation period of 26.5 minutes. The observations were part of a controlled exposure experiment designed to study how dolphins respond to naval sonar. In this experiment, an attenuated sonar source was introduced at a distance of approximately 1.5 km from the dolphin group, about 14 minutes after the start of the observations. The sonar emitted pings every 25 seconds for a 10 minute exposure period. Following the end of the sonar exposure, dolphin movements were observed for an additional 2.7 minutes. Overall, the dataset captured dolphin movement before, during, and after the sonar exposure, providing a valuable setting to study whether and how social behavior changes under disturbance. Further details on the experimental design, image collection, and spatial tracking of dolphins can be found in \citet{irsmdce}.

The total number of dolphins observed at any time varied between as few as $0$ to as many as $9$, with a median of $5$ individuals. The multi-label dataset had 101 unique labels, yielding $101-n_1$ random row vectors in the
identity assignment matrices $\bm{\mathrm{Z}}_{1:T}$, where $n_1$ is the number of individuals observed at the first time step. Although total population sizes are rarely known, we verified a population of $17$ dolphins through a painstaking, time-consuming review of the drone-based imagery. In this particular application, it was possible through careful scrutiny of the images by an expert, to uniquely identify all individual dolphins, which provided us with an opportunity to compare the marginal posterior to the posterior conditioned on the true $\bm{\mathrm{Z}}_{1:T}$ as we did in the simulation study. The primary scientific goal of our analysis was to quantify the evidence that sonar exposure affected the social behavior of the dolphin group during and after the exposure. 

\begin{figure}[htbp]
    \hspace{-6mm}
    \includegraphics[width=1.01\textwidth,height=\textheight,keepaspectratio]{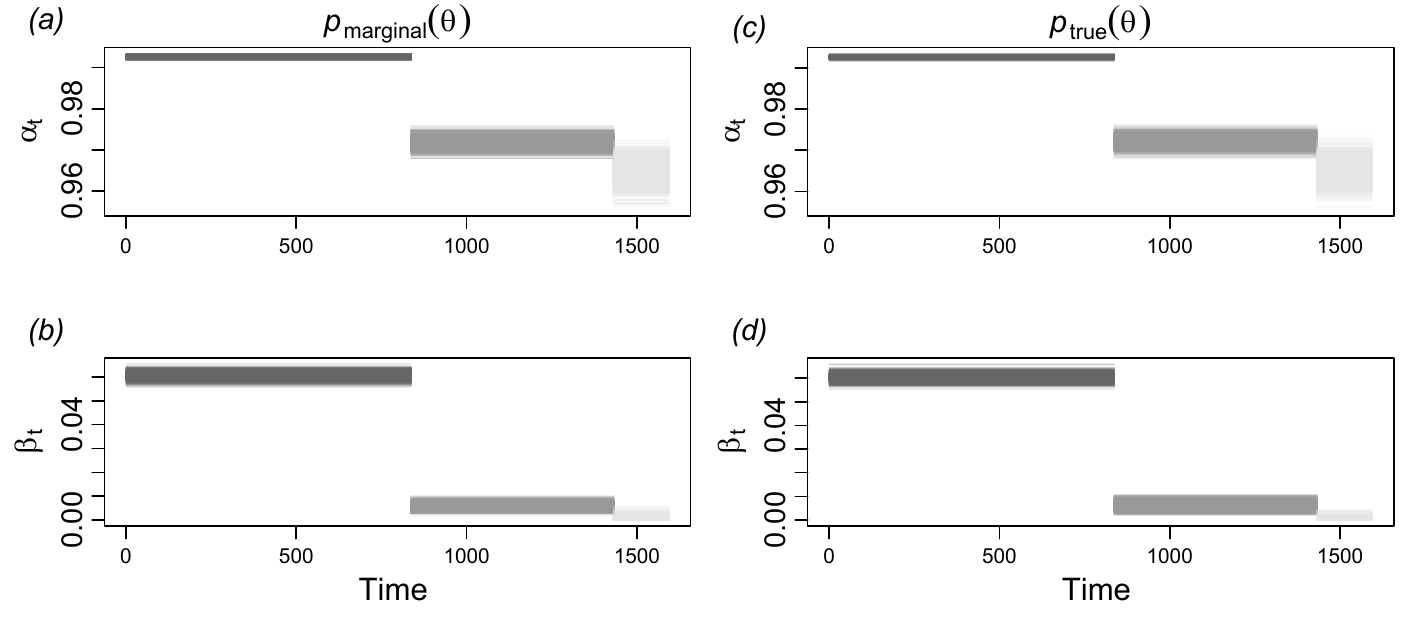}

\vspace{2mm}
    \caption{ Posterior summaries for alignment $( \alpha_t )$ and attraction $(\beta_t)$. Plots (a) and (b) correspond to the posteriors from $p_{\mathrm{marginal}}(\bm{\theta})$ and plot (c) and (d) correspond to $p_{\mathrm{true}}(\bm{\theta})$. Within each plot, the left, middle, and right segments represent the pre-exposure, exposure, and post-exposure phases of the sonar respectively.}

    \label{attr_align}
\end{figure}

To quantify how Risso's dolphin behavior changed during and after sonar exposure, we fitted the proposed model to the joint animal movement data $\bm{\mathrm{y}}_{1:T}$ with an unknown identity assignment matrix $\bm{\mathrm{Z}}_{1:T}$. We then evaluated these behavioral shifts by analyzing the posterior distributions of the regression coefficients to study $\alpha_{t}$ and $\beta_{t}$. Each grey horizontal line in Figure (\ref{attr_align}) represents a posterior sample of the behavioral parameter over time. The step changes occur at the transition points between the before, during, and after sonar-exposure periods, allowing the posterior distribution to reflect changes in social behavior across the three experimental phases. Similar to the simulation study, we work with model fits having distribution $p_{\mathrm{marginal}}(\bm{\theta})$, $p_{\mathrm{true}}(\bm{\theta})$, and $p_{\mathrm{random}}(\bm{\theta})$. 

Figure (\ref{attr_align}a) and (\ref{attr_align}b)  correspond to the posterior estimations of $\alpha_t$ and $\beta_t$ from $p_{\mathrm{marginal}}(\bm{\theta})$. For alignment, the posterior samples were high before exposure, roughly above $0.99$. During exposure, the posterior distribution shifted downward to approximately $0.97$, indicating reduced alignment. Following exposure, the posterior distribution shifted downward and became more diffuse, indicating that alignment remained reduced  and that the individuals were less synchronized with the socially connected individuals compared to their movement during the pre-exposure period. This increased posterior variation may be from the shorter time frame of the post-exposure observation period. For attraction, the pattern was similar to that observed for alignment. The overall attraction remained small throughout the models. Before exposure, the posterior was centered around approximately $0.06$, but during exposure it decreased to near zero. After exposure, it remained close to zero, suggesting that attraction was reduced during sonar exposure and did not recover afterward. The reduction in attraction and alignment suggests a disruption to typical social and movement behavior during sonar exposure. Individuals show less tendency to move towards one another or less tendency to align their movement in parallel to the socially connected individuals, indicating that individuals tend to have more individual based movement than group based movement. The reduction for the after exposure period suggested that the social behavior of the group may not recover immediately following the exposure. We observed a similar pattern in dolphin alignment to that reported by the authors of \citet{dynsn}; however, the attraction parameter showed a different pattern. \citet{dynsn} found a stronger attraction toward socially connected individuals during sonar exposure relative to the pre-exposure period.

 We compared the posterior estimations of the attraction and alignment parameters obtained from $p_{\mathrm{marginal}}(\bm{\theta})$  to those obtained from $p_{\mathrm{true}}(\bm{\theta})$ where we uniquely identify the individuals. We recognize that such validation with a known identity reference dataset may not be available in many applications, particularly for species where individuals cannot be reliably distinguished from images alone. We observed similar patterns of attraction and alignment across the three exposure phases in Figure (\ref{attr_align}c) and  (\ref{attr_align}d) compared to the posterior estimations from $p_{\mathrm{marginal}}(\bm{\theta})$. The consistency between the two posterior distributions is encouraging, as it suggests that the model can recover similar behavioral patterns even when individual identities are uncertain and inferred within the model. Thus, fitting the model with unknown identities may still provide reliable inference on attraction and alignment without requiring the identities of all individuals to be known with certainty.

 \begin{figure}[htbp]
    \centering
    \hspace{-3mm}\includegraphics[width=1.01\textwidth,height=\textheight, keepaspectratio]{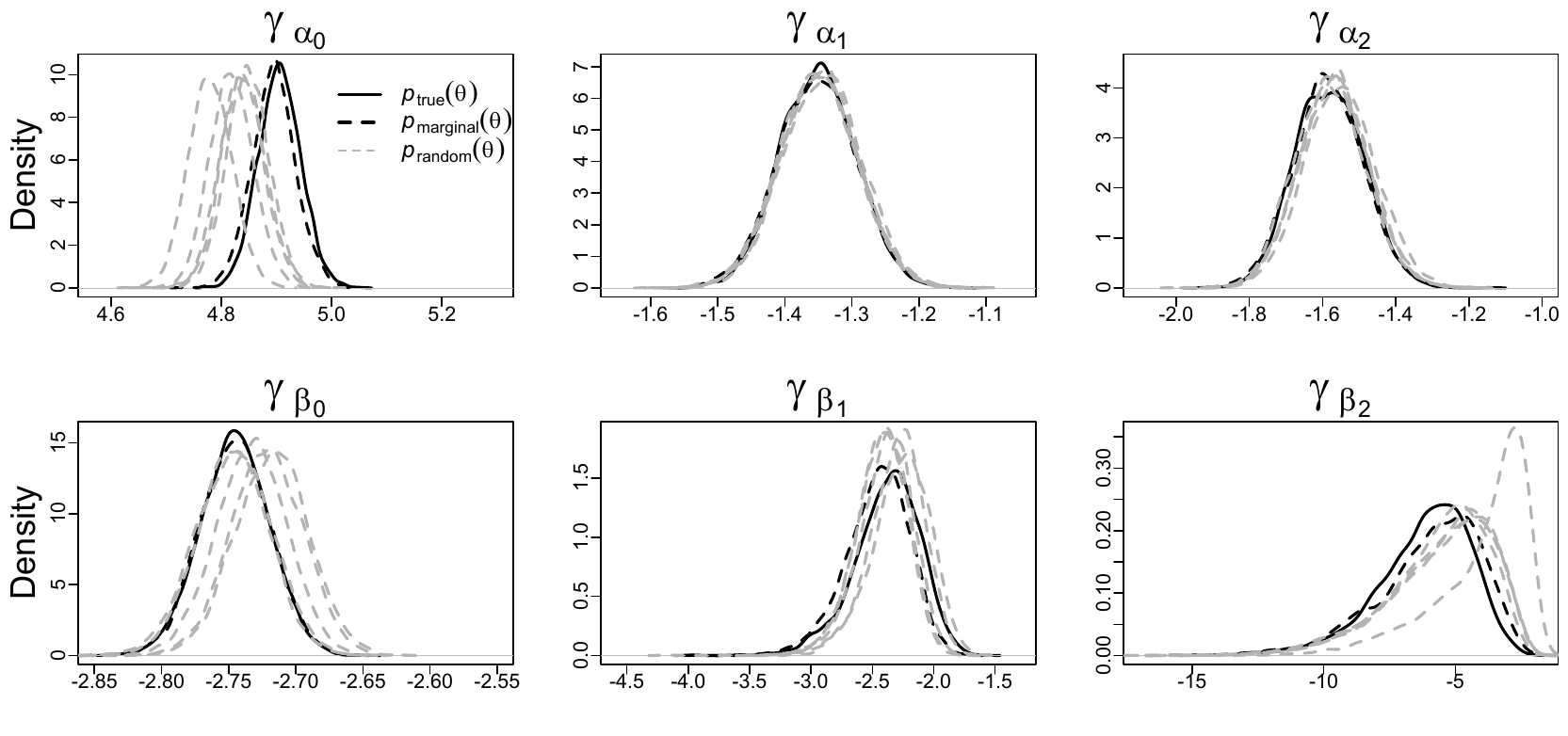}
    \vspace{2mm}
    \caption{Posterior density distributions of the regression coefficients under three model fits. The black line corresponds to the model fit when individual identities are correctly known, the dotted black line corresponds to the model fit that accounts for identity uncertainty and each dotted gray line corresponds to a different realizations of model fit when labels are assigned randomly to observations and then treated as fixed during model fit. }

    \label{dens_curve}
\end{figure}

Figure (\ref{dens_curve}) shows the posterior densities of the regression coefficients obtained from three different distributions. Across most components of $\bm{\gamma}_{\alpha}$ and $\bm{\gamma}_{\beta}$, the posterior density curves for these parameters obtained from $p_{\mathrm{marginal}}(\bm{\theta})$ closely overlap with those obtained from $p_{\mathrm{true}}(\bm{\theta})$. This similarity suggests that explicitly accounting for uncertainty in the identity assignments allows the model to recover posterior distributions for the behavioral regression coefficients that are comparable to those obtained when the true identities are available.  In contrast, the posterior estimation for  $\bm{\gamma}_{\alpha}$ and $\bm{\gamma}_{\beta}$ obtained from $p_{\mathrm{random}}(\bm{\theta})$ produces some shifts compared to the other two model fits, with noticeable deviations observed for $\gamma_{\alpha_0}$, $\gamma_{\beta_0}$ and $\gamma_{\beta_2}$. We notice similar patterns of deviation obtained from multiple fixed realizations of $\bm{\mathrm{Z}}_{1:T}^{\mathrm{random}}$. This shows that regression parameters are sensitive to identity assignments. If we assign random identities without accounting for this uncertainty, we risk getting misleading posterior estimations of their movement behavior.

\section{Discussion}
\label{sec:discussions}

The social lives of animals are hidden within their collective movements, which can provide insights into how they interact and behave as a group. Researchers have quantified these behaviors using summary statistics like average alignment, nearest-neighbor distance, and spatial group structures \citep{eldmagm,sosfdfs, dsnmag, dsnm,mifmia}. Our work examines how Risso's dolphins change their behavior when they encounter mid-frequency active sonar. Traditional tracking devices, such as GPS, radio signals, bio-loggers, and satellite networks can uniquely identify and monitor animals, but they need to be physically attached to the animals. Unmanned aerial vehicles (UAVs) are now being used by researchers to monitor these animals without physical contact \citep{irsmdce, brcdns}. While drones are powerful tools for collecting movement data to study and quantify behavior, they present the challenge of not being able to track the trajectories of individuals at all times.

We extended the Bayesian state-space models proposed by \citet{dsnm} and \citet{dynsn} for estimating behavioral parameters such as alignment and attraction, by treating the identity-assignment matrix as random. In our application, observations recorded at one-second intervals allowed us to use the available label information to impose recognition and exclusion constraints, reducing the dimensionality of  $\bm{\mathrm{Z}}_{1:T}$. Simulation and empirical analysis demonstrate that accounting for identity uncertainty yields posterior estimates of movement parameters comparable to those obtained when individuals are uniquely identified, whereas assigning incorrect fixed identities can lead to distorted estimates.

 A natural extension would be to incorporate information from a dynamic social network $\bm{\mathrm{W}}_{1:T}$, as defined by authors \citet{dsnm}, rather than treating the network as fixed and fully connected. Individuals that are repeatedly observed together or exhibit similar movement patterns may be more likely to remain socially connected, and this information could be used to construct more informative prior probabilities for the identity assignments. Such information would not necessarily reduce the number of random rows in $\bm{\mathrm{Z}}_{1:T}$, but could concentrate posterior probability on assignments that are more consistent with the inferred social structure.

One important limitation of our method is the assumption that $J$ is fixed and known. We made this assumption due to being aware of the total individuals that were available during the study period, however in other applications, this may not be known. Future work could treat $J$ as unknown and infer it jointly with the other unknown parameters in the model. However, doing so would introduce additional complexity because the number of individuals represented in the model, and
consequently the dimension of the parameter space, could vary across MCMC iterations. Such an extension could be implemented using an
alternative sampling framework, such as reversible-jump
MCMC, to allow transitions between parameter spaces
of different dimensions.

\subsection{Additional future extensions}
    
    Although treating the identity-assignment matrices $\bm{\mathrm{Z}}_{1:T}$ as unknown random variables is a central component of our proposed framework, the primary objective of this study is inference on the movement parameters while propagating uncertainty in individual identity. A natural direction for future work, particularly for researchers interested in individual identification, is to study the posterior distribution of $\bm{\mathrm{Z}}_{1:T}$ in detail. Posterior samples of $\bm{\mathrm{Z}}_{1:T}$ could, for example, also be used to estimate the probability that two disconnected trajectory segments belong to the same underlying individual by calculating the proportion of posterior samples in which the two segments are assigned to the same posterior label. Because true identities are rarely known in real-world applications, these posterior matching probabilities generally cannot be directly verified if they come from the same individual. Their accuracy could instead be evaluated through simulation studies, where the true identities are known. Due to the possibility of label switching, validation should not require the inferred reference labels to match the labels used to generate the simulated data. Instead, validation should assess whether two trajectory segments assigned to the same identity in the posterior belong to the same underlying individual.

\section*{Acknowledgments}

 This project was supported in part by the US Office of Naval Research, Marine Mammals and Biology Program (awards N000141912572 and N000142312497 to B. Southall (principal investigator)).

\textbf{Data Availability Statement}: The authors confirm that the data and replication code supporting this study's results are contained within the article's supplementary files. These materials will be provided upon the manuscript's acceptance.
\bibliographystyle{apalike} 
\bibliography{movement}

@article{dynsn,
  title={Detecting changes in dynamic social networks using multiply-labeled movement data},
  author={Boulil, Zaineb L and Durban, John W and Fearnbach, Holly and Joyce, Trevor W and Leander, Samantha GM and Scharf, Henry R},
  journal={Journal of Agricultural, Biological and Environmental Statistics},
  volume={28},
  number={2},
  pages={243--259},
  year={2023},
  publisher={Springer}
}

@article{dsnm,
  title={Dynamic social networks based on movement},
  author={Scharf, Henry R. and Hooten, Mevin B. and Fosdick, Bailey K. and Johnson, Devin S. and London, Josh M. and Durban, John W.},
  journal={The Annals of Applied Statistics},
  volume={10},
  number={4},
  pages={2182--2202},
  year={2016},
  publisher={Institute of Mathematical Statistics}
}

@article{brcdns,
  title={Behavioural responses of common dolphins to naval sonar},
  author={Southall, Brandon L and Durban, John W and Calambokidis, John and Casey, Caroline and Fahlbusch, James A and Fearnbach, Holly and Flynn, Kiirsten R and Fregosi, Selene and Friedlaender, Ari S and Leander, Samantha GM and others},
  journal={Royal Society Open Science},
  volume={11},
  number={10},
  pages={240650},
  year={2024},
  publisher={The Royal Society}
}

@article{eldmagm,
  title={Effective leadership and decision-making in animal groups on the move},
  author={Couzin, Iain D and Krause, Jens and Franks, Nigel R and Levin, Simon A},
  journal={Nature},
  volume={433},
  number={7025},
  pages={513--516},
  year={2005},
  publisher={Nature Publishing Group UK London}
}

@manual{nimble2026,
  title = {NIMBLE: MCMC, Particle Filtering, and Programmable Hierarchical Modeling},
  author = {de Valpine, Perry and Paciorek, Christopher and Turek, Daniel and Michaud, Nicholas and Anderson-Bergman, Christian and Obermeyer, Fritz and Wehrhahn Cortes, Claudia and Rodr{\'i}guez, Adriana and Temple Lang, Duncan and Paganin, Stefano},
  year = {2026},
  note = {R package version 1.4.1},
  doi = {10.5281/zenodo.1211190}
}

@book{tsassm,
  title={Time series analysis by state space methods},
  author={Durbin, James and Koopman, Siem Jan},
  year={2012},
  publisher={Oxford University Press (UK)}
}

@article{srd,
  title={Social structure of Risso’s dolphins (Grampus griseus) at the Azores: a stratified community based on highly associated social units},
  author={Hartman, Karin L and Visser, Fleur and Hendriks, Arthur JE},
  journal={Canadian Journal of Zoology},
  volume={86},
  number={4},
  pages={294--306},
  year={2008}
}

@article{irsmdce,
  title={Integrating remote sensing methods during controlled exposure experiments to quantify group responses of dolphins to navy sonar},
  author={Durban, JW and Southall, BL and Calambokidis, J and Casey, C and Fearnbach, H and Joyce, TW and Fahlbusch, JA and Oudejans, MG and Fregosi, S and Friedlaender, AS and others},
  journal={Marine Pollution Bulletin},
  volume={174},
  pages={113194},
  year={2022},
  publisher={Elsevier}
}

@article{mgdam,
  title={Modelling group dynamic animal movement},
  author={Langrock, Roland and Hopcraft, J Grant C and Blackwell, Paul G and Goodall, Victoria and King, Ruth and Niu, Mu and Patterson, Toby A and Pedersen, Martin W and Skarin, Anna and Schick, Robert S},
  journal={Methods in Ecology and Evolution},
  volume={5},
  number={2},
  pages={190--199},
  year={2014},
  publisher={Wiley Online Library}
}

@article{mifmia,
  title={Modelling and inference for the movement of interacting animals},
  author={Milner, Jordan E and Blackwell, Paul G and Niu, Mu},
  journal={Methods in Ecology and Evolution},
  volume={12},
  number={1},
  pages={54--69},
  year={2021},
  publisher={Wiley Online Library}
}

@article{dsnmag,
  title={Distinguishing social from nonsocial navigation in moving animal groups},
  author={Bode, Nikolai WF and Franks, Daniel W and Wood, A Jamie and Piercy, Julius JB and Croft, Darren P and Codling, Edward A},
  journal={The American Naturalist},
  volume={179},
  number={5},
  pages={621--632},
  year={2012},
  publisher={University of Chicago Press Chicago, IL}
}

@article{amsqi,
  title={A video multitracking system for quantification of individual behavior in a large fish shoal: advantages and limits},
  author={Delcourt, Johann and Becco, Christophe and Vandewalle, Nicolas and Poncin, Pascal},
  journal={Behavior research methods},
  volume={41},
  number={1},
  pages={228--235},
  year={2009},
  publisher={Springer}
}

@article{idtiga,
  title={idTracker: tracking individuals in a group by automatic identification of unmarked animals},
  author={P{\'e}rez-Escudero, Alfonso and Vicente-Page, Juli{\'a}n and Hinz, Robert C and Arganda, Sara and De Polavieja, Gonzalo G},
  journal={Nature methods},
  volume={11},
  number={7},
  pages={743--748},
  year={2014},
  publisher={Nature Publishing Group US New York}
}

@article{taeatowtma,
  title={ToxId: an efficient algorithm to solve occlusions when tracking multiple animals},
  author={Rodriguez, Alvaro and Zhang, Hanqing and Klaminder, Jonatan and Brodin, Tomas and Andersson, Magnus},
  journal={Scientific reports},
  volume={7},
  number={1},
  pages={14774},
  year={2017},
  publisher={Nature Publishing Group UK London}
}

@article{mape,
  title   = {Multi-animal pose estimation, identification and tracking with {DeepLabCut}},
  author  = {Lauer, Jessy and
             Zhou, Mu and
             Ye, Shaokai and
             Menegas, William and
             Schneider, Steffen and
             Nath, Tanmay and
             Rahman, Mohammed Mostafizur and
             Di Santo, Valentina and
             Soberanes, Daniel and
             Feng, Guoping and
             Murthy, Venkatesh N. and
             Lauder, George and
             Dulac, Catherine and
             Mathis, Mackenzie Weygandt and
             Mathis, Alexander},
  journal = {Nature Methods},
  volume  = {19},
  number  = {4},
  pages   = {496--504},
  year    = {2022}
}

@article{irapsn,
  title={Identifying the role that animals play in their social networks},
  author={Lusseau, David and Newman, Mark EJ},
  journal={Proceedings of the Royal Society of London. Series B: Biological Sciences},
  volume={271},
  number={Suppl 6},
  pages={S477--S481},
  year={2004},
  publisher={The Royal Society}
}

@article{mgbsic,
  title={Modelling group movement with behaviour switching in continuous time},
  author={Niu, Mu and Frost, Fay and Milner, Jordan E and Skarin, Anna and Blackwell, Paul G},
  journal={Biometrics},
  volume={78},
  number={1},
  pages={286--299},
  year={2022},
  publisher={Wiley Online Library}
}

@article{ammfmi,
  title={Animal movement models for multiple individuals},
  author={Scharf, Henry R and Buderman, Frances E},
  journal={Wiley Interdisciplinary Reviews: Computational Statistics},
  volume={12},
  number={6},
  pages={e1506},
  year={2020},
  publisher={Wiley Online Library}
}

@article{jemibsiamd,
  title={Joint estimation over multiple individuals improves behavioural state inference from animal movement data},
  author={Jonsen, Ian},
  journal={Scientific reports},
  volume={6},
  number={1},
  pages={20625},
  year={2016},
  publisher={Nature Publishing Group UK London}
}

@article{dpvafuas,
  title={Determining position, velocity and acceleration of free-ranging animals with a low-cost unmanned aerial system},
  author={Harvey, Richard J and Roskilly, Kyle and Buse, Chris and Evans, Hannah K and Hubel, Tatjana Y and Wilson, Alan M},
  journal={Journal of Experimental Biology},
  volume={219},
  number={17},
  pages={2687--2692},
  year={2016},
  publisher={The Company of Biologists Ltd}
}

@article{dbhrtav,
  title={Drone-based high-resolution tracking of aquatic vertebrates},
  author={Raoult, Vincent and Tosetto, Louise and Williamson, Jane E},
  journal={Drones},
  volume={2},
  number={4},
  pages={37},
  year={2018},
  publisher={MDPI}
}

@article{irsimc,
  title={Inferring the rules of social interaction in migrating caribou},
  author={Torney, Colin J and Lamont, Myles and Debell, Leon and Angohiatok, Ryan J and Leclerc, Lisa-Marie and Berdahl, Andrew M},
  journal={Philosophical Transactions of the Royal Society B: Biological Sciences},
  volume={373},
  number={1746},
  pages={20170385},
  year={2018}
}

@article{qmbec,
  title={Quantifying the movement, behaviour and environmental context of group-living animals using drones and computer vision},
  author={Koger, Benjamin and Deshpande, Adwait and Kerby, Jeffrey T and Graving, Jacob M and Costelloe, Blair R and Couzin, Iain D},
  journal={Journal of Animal Ecology},
  volume={92},
  number={7},
  pages={1357--1371},
  year={2023},
  publisher={Wiley Online Library}
}

@article{bwrsans,
  title={Beaked whales respond to simulated and actual navy sonar},
  author={Tyack, Peter L and Zimmer, Walter MX and Moretti, David and Southall, Brandon L and Claridge, Diane E and Durban, John W and Clark, Christopher W and D'amico, Angela and DiMarzio, Nancy and Jarvis, Susan and others},
  journal={PloS one},
  volume={6},
  number={3},
  pages={e17009},
  year={2011},
  publisher={Public Library of Science San Francisco, USA}
}

@article{ammhmb,
  title={A multivariate mixed hidden Markov model for blue whale behaviour and responses to sound exposure},
  author={DeRuiter, Stacy L and Langrock, Roland and Skirbutas, Tomas and Goldbogen, Jeremy A and Calambokidis, John and Friedlaender, Ari S and Southall, Brandon L},
  journal={The Annals of Applied Statistics},
   volume={11},
  number={1},
  pages={362-392},
  year={2017}
}

@article{brstbbw,
  title={Behavioral responses of satellite tracked Blainville's beaked whales (Mesoplodon densirostris) to mid-frequency active sonar},
  author={Joyce, Trevor W and Durban, John W and Claridge, Diane E and Dunn, Charlotte A and Hickmott, Leigh S and Fearnbach, Holly and Dolan, Karin and Moretti, David},
  journal={Marine Mammal Science},
  volume={36},
  number={1},
  pages={29--46},
  year={2020},
  publisher={Wiley Online Library}
}

@article{nbwipers,
  title={Northern bottlenose whales in a pristine environment respond strongly to close and distant navy sonar signals},
  author={Wensveen, Paul J and Isojunno, Saana and Hansen, Rune R and von Benda-Beckmann, Alexander M and Kleivane, Lars and Van Ijsselmuide, Sander and Lam, Frans Peter A and Kvadsheim, Petter H and DeRuiter, Stacy L and Cur{\'e}, Charlotte and others},
  journal={Proceedings of the Royal Society B: Biological Sciences},
  volume={286},
  number={1899},
  pages={20182592},
  year={2019}
}

@article{sosfdfs,
  title={Self-organized shape and frontal density of fish schools},
  author={Hemelrijk, Charlotte K and Hildenbrandt, Hanno},
  journal={Ethology},
  volume={114},
  number={3},
  pages={245--254},
  year={2008},
  publisher={Wiley Online Library}
}

@article{scrpi,
  title={Spatial capture--recapture with partial identity: An application to camera traps},
  author={Augustine, Ben C and Royle, J Andrew and Kelly, Marcella J and Satter, Christopher B and Alonso, Robert S and Boydston, Erin E and Crooks, Kevin R},
   journal={The Annals of Applied Statistics},
  volume={12},
  number={1},
  pages={67--95},
  year={2018}
}
\end{document}